\documentclass[
 reprint,
 superscriptaddress,
 amsmath,
 amssymb,
 aps,
 prb,
 citeautoscript,
 floatfix,
 longbibliography
]{revtex4-2}

\usepackage{graphicx}
\usepackage{dcolumn}
\usepackage{bm}
\usepackage{physics2}
\usephysicsmodule{ab,braket}
\usepackage{url}
\usepackage{makecell}
\usepackage{hyperref}
\hypersetup{
  colorlinks   = true,    
  urlcolor     = blue,    
  linkcolor    = blue,    
  citecolor    = blue     
}
\begin{document}

\preprint{APS/123-QED}

\title{Fast microwave-driven two-qubit gates\\between fluxonium qubits with a transmon coupler }

\author{Siddharth Singh}
\thanks{these authors contributed equally}
\email{Siddharth.Singh@tudelft.nl}
\affiliation{QuTech and Kavli Institute of Nanoscience, Delft University of Technology, 2628 CJ, Delft, The Netherlands}

\author{Eugene Y. Huang}
\thanks{these authors contributed equally}
\email{E.Y.Huang@tudelft.nl}
\affiliation{QuTech and Kavli Institute of Nanoscience, Delft University of Technology, 2628 CJ, Delft, The Netherlands}

\author{Jinlun Hu}
\affiliation{QuTech and Kavli Institute of Nanoscience, Delft University of Technology, 2628 CJ, Delft, The Netherlands}

\author{Figen Yilmaz}
\affiliation{QuTech and Kavli Institute of Nanoscience, Delft University of Technology, 2628 CJ, Delft, The Netherlands}

\author{Martijn F.S. Zwanenburg}
\affiliation{QuTech and Kavli Institute of Nanoscience, Delft University of Technology, 2628 CJ, Delft, The Netherlands}

\author{Piranavan Kumaravadivel}
\affiliation{Netherlands Organization for Applied Scientific Research (TNO), Delft, The Netherlands}

\author{Siyu Wang}
\affiliation{QuTech and Kavli Institute of Nanoscience, Delft University of Technology, 2628 CJ, Delft, The Netherlands}

\author{Taryn V. Stefanski}
\thanks{Present address: QphoX, 2628 XG, Delft, The Netherlands}
\affiliation{QuTech and Kavli Institute of Nanoscience, Delft University of Technology, 2628 CJ, Delft, The Netherlands}
\affiliation{Quantum Engineering Centre for Doctoral Training, H. H. Wills Physics Laboratory and Department of Electrical and Electronic Engineering, University of Bristol, BS8 1FD, Bristol, UK}

\author{Christian Kraglund Andersen}
\email{C.K.Andersen@tudelft.nl}
\affiliation{QuTech and Kavli Institute of Nanoscience, Delft University of Technology, 2628 CJ, Delft, The Netherlands}

\begin{abstract}
Two qubit gates constitute fundamental building blocks in the realization of large-scale quantum devices. Using superconducting circuits, two-qubit gates have previously been implemented in different ways with each method aiming to maximize gate fidelity. Another important goal of a new gate scheme is to minimize the complexity of gate calibration. In this work, we demonstrate a high-fidelity two-qubit gate between two fluxonium qubits enabled by an intermediate capacitively coupled transmon. The coupling strengths between the qubits and the coupler are designed to minimize residual crosstalk while still allowing for fast gate operations. The gate is based on frequency selectively exciting the coupler using a microwave drive to complete a 2$\pi$ rotation, conditional on the state of the fluxonium qubits. When successful, this drive scheme implements a conditional phase gate. Using analytically derived pulse shapes, we minimize unwanted excitations of the coupler and obtain gate errors of $10^{-2}$ for gate times below 60~ns. At longer durations, our gate is limited by relaxation of the coupler. Our results show how carefully designed control pulses can speed up frequency selective entangling gates.
\end{abstract}
\maketitle

\section{\label{sec:level1}Introduction}
Superconducting qubits are a leading candidate for building large quantum processors in recent years, demonstrating significant progress towards quantum simulations~\cite{Googlesim,siddiqisimulation,IBMsimulation,Nanjingsim,paintersim,kirchmairsim,rosen2024flat,Rosen2024} and quantum error correction~\cite{acharya2024quantum,Krinner2022,Marques2022,putterman2024hardwareefficientquantumerrorcorrection}. Within these experiments, single and two-qubit gates are fundamental building blocks that dictate overall  performance. The most widespread type of qubit used in these applications is the transmon qubit~\cite{Koch2007}. Two-qubit gates for transmon qubits were originally implemented with high fidelities using dynamic flux pulses to tune the qubits in and out of resonance~\cite{barends2019diabatic,DiCarlo2009,Dynamicfluxpulse} or by applying a strong microwave drive to activate an effective two-qubit interaction~\cite{Chow_2013,Chow2011,mitchell2021hardware,sheldon2016procedure,134507}. However, residual crosstalk often limits such architectures~\cite{krinner2020benchmarking,sundaresan2020reducing}. More recently, tunable couplers have consistently reached high gate fidelities alleviating the crosstalk issue to a great extent~\cite{collodo2020implementation,ye2021realization,marxer2023long,Li2018,Xu2021,sete2021parametric,Arute2019}. 

Despite transmon based gates reaching higher fidelities, transmons have a small anharmonicity which eventually limits the speed of gates and can lead to leakage out of the computational subspace~\cite{miao2023overcoming,LRU,werninghaus2021leakage}. The small anharmonicity further imposes strong requirements on qubit frequency targeting to avoid frequency crowding~\cite{IBMcollision,morvan2022optimizing}. Additionally, dielectric losses are known to limit the coherence times of transmons~\cite{Ganjam2024,Read2023}.

The fluxonium qubit is one appealing alternative which aim to mitigate these shortcomings as it is less prone to dielectric loss because of its lower qubit frequency and while also featuring a much larger anharmonicity~\cite{manucharyan2009fluxonium,nguyen2019high}.
Coherence times in the millisecond range have been demonstrated in recent years~\cite{somoroff2023millisecond,wang2024achieving,earnest2018realization}.
Fast single-qubit and two-qubit gates have also been shown with high fidelities~\cite{zwanenburg2025,ding2023high,quentin2021}. One challenge in implementing fast two-qubit gates between fluxonium qubits comes from their small charge dipole moment, suppressing the interaction between the computational states when directly capacitively coupled. Thus, direct capacitive coupling schemes for fluxonium qubits may require very large capacitors or very strong driving~\cite{dogan2023two,quentin2021}; however, stronger coupling schemes are often accompanied by an increased residual interaction rate.
A possible remedy is to use inductive coupling~\cite{zhang2024tunable,ma2024native}. However, inductive coupling may be accompanied with additional flux loops and require closer placement of qubits. Similarly to architectures for transmon qubits, an apparent solution is to introduce a tunable coupler to mediate strong interaction between the fluxonium qubits while simultaneously suppressing any residual coupling between the computational states~\cite{ding2023high,zhang2024tunable,simakov2023coupler}. In particular, two-qubit gates with high fidelities have been implemented by applying a microwave drive to the coupler~\cite{ding2023high,moskalenko2022high}. While high fidelities have been achieved with this gate scheme, the maximum reported fidelity utilized a reinforcement learning algorithm for gate calibration~\cite{ding2023high}, a potentially time costly procedure. Moreover, while reinforcement learning algorithms can adapt to unknown situations, these black-box algorithms often lead to calibrated pulse shapes which are hard to interpret. 

\begin{figure}
    \centering
    \includegraphics[width=1\linewidth]{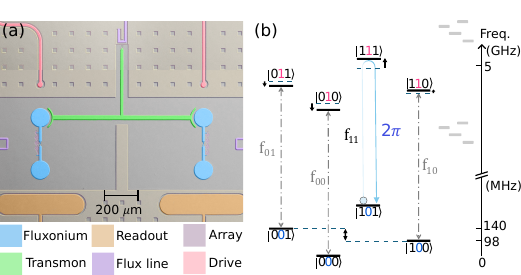}
    \caption{(a) False colored microscope image of the device zoomed in to a single set of the combined qubit system. The fluxonium qubits are indicated in blue and are coupled by a transmon displayed in green. All the qubits are flux biased by currents applied to the flux lines (shown in purple), and the fluxonium qubits are driven capacitively by the drive lines (shown in pink). The readout resonators (shown in orange) have large capacitive elements near the qubits and beyond this image, the resonators extend into $\lambda/2$ resonators.
(b) Energy diagram of the combined system with different levels. We use the notation $|F_1 C F_2\rangle $ to indicate the state of fluxonium 1, coupler and fluxonium 2, respectively. For clarity, we mark the system states corresponding to the coupler ground state in blue, constituting the computational subspace. The states with the coupler in the excited state are represented in red. We indicate the transition frequencies of the coupler with $f_{ij} = \omega_{ij}/(2\pi)$ conditioned on the fluxonium qubits' state $|ij\rangle$. Our gate scheme is operated by driving a $2\pi$ rotation on the coupler when the fluxonium qubits are in the $|11\rangle$ state indicated with the blue array. The gray energy levels on the right-hand side indicate the higher excited fluxonium states. These states shift the dressed coupler levels (bold black levels) away from the bare coupler levels (dash-dotted levels) allowing for each coupler transition to be selectively addressed.}
    \label{fig:1}
\end{figure}

In this work, we implement a two-qubit gate between two fluxonium qubits coupled via a transmon coupler circuit, see Fig.~\ref{fig:1}(a). The gate is implemented by driving a coupler transition that is conditional on the state of the two fluxonium qubits, see Fig.~\ref{fig:1}(b). The transmon coupler interacts strongly with the higher levels of the fluxonium qubits which lead to distinct coupler transition frequencies for each computational state. We expect that a large frequency selectivity among these transitions lead to faster two-qubit gates. On the other hand, for faster gates, the spectral overlap between the pulse and unwanted coupler transitions lead to coherent errors in the gate. Here, we employ simple analytical pulse shaping techniques~\cite{motzoi2009simple,hyyppa2024reducing} to achieve fast two-qubit gates. Given the analytical form of these pulse shapes, we can interpret how they minimize coherent errors. Specifically, the pulses are designed to minimize Fourier amplitude at the unwanted transitions. We experimentally obtain two-qubit gate fidelities of 99\% for gate times as low as 60~ns.

\section{\label{sec:device description}Device description}

The device used in this work is composed of two fluxonium qubits, see Fig.~\ref{fig:1}(a). These two fluxonium qubits are both capacitively coupled to a transmon coupler circuit in a geometry that accommodates up to four couplers connected to each fluxonium qubit as required e.g. for implementing the surface code~\cite{Fowler_surface,Versluis,andersen2020repeated,google2023suppressing}. Throughout this work, we refer to the two fluxonium qubits as $F_1$ and $F_2$ and to the transmon coupler as $C$. The fluxonium qubits and the transmon couplers feature Josephson junctions fabricated using a Manhattan style process, see further details about the fabrication in Appendix~\ref{sec:fabrication} and in Ref.~\cite{yilmaz2024energy}. The fluxonium qubits include a flux loop each composed of the central Josephson junction with Josephson energy $E_J$ and a Josephson array consisting of 100 junctions in series with a combined inductive energy of $E_L$. The flux loops of the fluxonium qubits as well as the symmetric SQUID-loop of the transmon coupler can be biased using their individual flux lines. Note that the coupler circuit is also coupled weakly to each drive line, such that the coupler can be readily driven by either. The fluxonium qubits are read out using dispersive readout enabled by their individual readout resonators.
Similarly, the transmon coupler is also coupled to each readout resonator. The parameters of the device are further described in Appendix~\ref{sec:Design parameters}.

We write the total system Hamiltonian as
\begin{equation}\label{eq:1}
    \hat{H} = \hat{H}_{F_1} + \hat{H}_{F_2} + \hat{H}_{C} + \hat{H}_{I} 
\end{equation}
where each fluxonium qubit is described by their individual Hamiltonian
\begin{equation}\label{eq:2}
    \hat{H}_{F_i} = 4E_{Ci}\hat{n}_{F_i}^2 + \frac{1}{2}E_{Li}\hat{\phi}_{F_i}^2 - E_{Ji} \cos(\hat{\phi}_{F_i} - \phi_{i,ext}),
\end{equation}
where $E_{Ci}$ is the charging energy,$E_{Li}$ is the inductive energy and $E_{Ji}$ is the Josephson energy  of the $i$th fluxonium and $\phi_{i,ext}$ is the external flux-bias of each qubit. Similarly, the Hamiltonian for the transmon coupler is
\begin{equation}\label{eq:3}
     \hat{H}_{C}= 4E_{C,C}\hat{n}_{C}^2 + E_{J,C}|\cos(\phi_{C,ext})|\,\cos(\hat{\phi}_C),
\end{equation}
where $E_{C,C}$ is the charging energy of the coupler and $E_{J,C}$ is the total Josephson energy of the SQUID. The transmon is tuned by the external flux $\phi_{C,ext} = \pi \Phi_{C,ext}/\Phi_0$ where $\Phi_0$ is the magnetic flux quantum. Since we have capacitive coupling between each fluxonium and the coupler as well as directly between the two fluxonium qubits, the interaction Hamiltonian is
\begin{equation}\label{eq:4}
        \hat{H_{I}} = \hbar g_{12}\hat{n}_{F_1}\hat{n}_{F_2}+\sum_{n=1,2}\hbar g_{ic}\hat{n}_{F_i}\hat{n}_{C} 
\end{equation}
We describe the quantum state of this coupled system using the basis state $\ket{F_1 C F_2}$, referring to the eigenstates of the coupled system unless otherwise mentioned. As indicated in Fig.~\ref{fig:1}(b), the higher excited states of each fluxonium qubit couple strongly with the transmon coupler leading to level repulsions between the transmon levels and the fluxonium levels. In particular, the transition frequency of the coupler will depend on the state of the fluxonium qubits. We will refer to the transition frequencies of the coupler as $\omega_{ij} = (E_{\ket{i1j}} - E_{\ket{i0j}})/\hbar$ where $E_{\ket{x}}$ is the eigen-energy of state $\ket{x}$. Since $\omega_{11} \neq \omega_{00}, \omega_{01}, \omega_{10}$, we can drive a selective $2\pi$ rotation on the $\omega_{11}$ transition such that the state $\ket{101}$ picks up a phase of $\pi$ and thereby implementing a conditional phase gate. 

\begin{figure}[t]
    \centering
    \includegraphics[width=\linewidth]{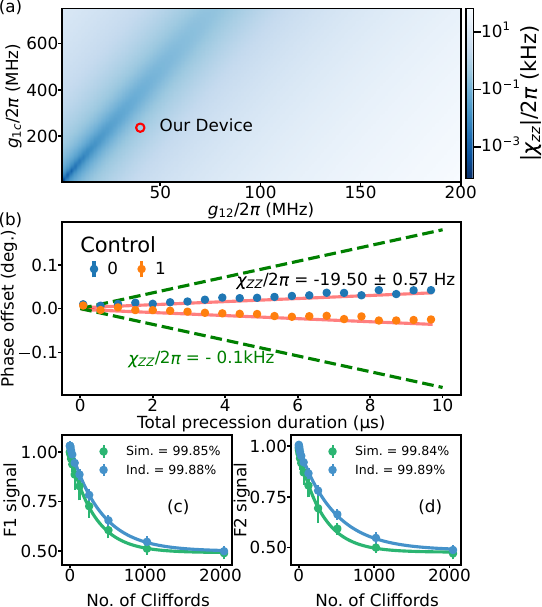}
    \caption{(a) The residual ZZ rate for a range of coupling strengths between the two fluxonium qubits, $g_{12}$, and between each fluxonium and the coupler, $g_{ic}$. Parameter values corresponding to our device are indicated in red circle.
    (b) Measured accumulated phase on $F_2$ dependent on the state of $F_1$. We extract the residual ZZ rate, $\chi_{ZZ}$, from the slope of the signal. For reference, the theoretical value of $\chi_{ZZ} = 0.1$~kHz is plotted in green. 
    (c, d) Randomized benchmarking for single qubit gates on fluxonium qubit $F_1$ and $F_2$, respectively, performed on each individual qubit (blue) and simultaneously on both qubits (green). The inset text displays the fitted average fidelity per gate.}
    \label{fig:2}
\end{figure} 
The specific fluxonium qubits that we study in this work are operated with qubit frequencies at 98 and 140~MHz for $F_1$ and $F_2$, respectively. The transmon coupler has a transition frequency around 4.7~GHz when operated at zero flux bias and with the fluxonium qubits in their ground state. With these qubit frequencies, the device is in a regime where the residual $ZZ$ coupling between is predicted to be below 0.1~kHz, see Fig.~\ref{fig:2}(a) where we extract the theoretical $ZZ$-coupling based on a numerical diagonalization of the full Hamiltonian. Here, we define the residual $ZZ$ coupling as
\begin{equation}\label{eq:residual-zz-definition}
       \chi_{ZZ} = \omega_{101} -\omega_{100} -\omega_{001} +\omega_{000}.
\end{equation}
To verify this residual coupling experimentally, we perform a modified Ramsey sequence to precisely measure the accumulated phase of $F_2$ with $F_1$ in either state $\ket{0}$ or in $\ket{1}$, see also Appendix~\ref{zz}. As seen in Fig.~\ref{fig:2}(b), we extract a residual $ZZ$ coupling of around $-20$~Hz, well below 0.1~kHz in magnitude. Thus, we expect the residual $ZZ$ coupling to cause little correlated crosstalk errors.

To further characterize the crosstalk between the two qubits, we employ simultaneous randomized benchmarking~\cite{Gambetta_RB, silva2024hands}, see Fig.~\ref{fig:2}(c) and (d). For these results, we decomposed the Clifford group into combinations of $\pi/2$-pulses and virtual-$Z$ gates~\cite{Virtual-z-gates}. Note that the readout signal is slightly shifted for the simultaneous protocol due to a shift in the readout signal arising from the microwave reset scheme of the fluxonium qubits, see also Appendix~\ref{reset}. We observe that the single-qubit gate fidelity is only slightly affected by simultaneous driving. Due to the small residual $ZZ$ coupling, we attribute the small degradation in the single qubit fidelities to microwave crosstalk between the fluxonium qubits.

\begin{figure}
    \centering
    \includegraphics[width=1\linewidth]{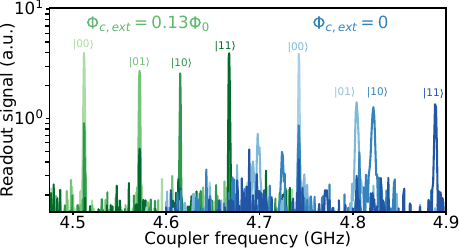}
    \caption{Conditional spectroscopy of the coupler state when fluxoniums $F_1,F_2$ are prepared in different states~$|F_1,F_2\rangle$, labeled above each peak. Two sets of peaks are observed for different flux operating points of the coupler, corresponding to $\Phi_{c,ext} =0 $ shown in shades of blue and $\Phi_{c,ext} = 0.13 \Phi_0$ shown in shades of green.} 
    \label{fig:3}
\end{figure}

\begin{figure}    \includegraphics[width=1\linewidth]{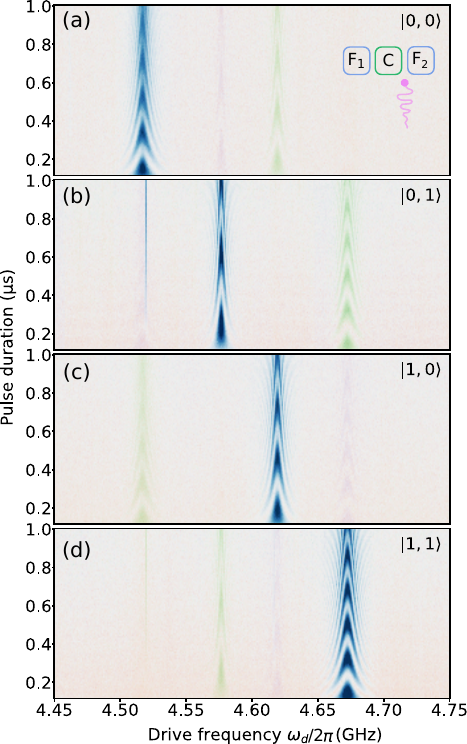}
    \caption{Rabi oscillations when driving near the coupler frequencies are shown with different initialization of the fluxonium state, $\ket{00}$, $\ket{01}$, $\ket{10}$ and $\ket{11}$ in panels (a), (b), (c) and (d), respectively. The readout signal of both fluxonium qubits are analyzed using principal component analysis and we visualize the main components in blue-to-green and red-to-purple. The coupler is driven through the drive line of $F_2$ as indicated in the inset of (a). }
    \label{fig:4}
\end{figure}
 
\section{\label{sec:Implementation}Two-qubit gate implementation}
Our gate scheme involves driving the microwave transitions of the transmon coupler. To verify that we can selectively drive the coupler based on the state of the fluxonium qubits, we perform conditional spectroscopy of the coupler, see also Appendix~\ref{Coupler readout}. As shown in Fig. \ref{fig:3}, we see four distinct frequency transitions between 4.5~GHz and 4.9~GHz for two different flux bias points. Due to fabrication uncertainty, the coupler frequency in the device lies close to the resonance frequency of the readout resonator of $F_1$ at $4.993$~GHz. As a consequence, the coupler is partly hybridized with the readout resonator and we observe an overall degradation of qubit coherence. To mitigate this detrimental effect, we use the frequency flexible aspect of our design and flux bias the coupler to $0.13\Phi_0$ for the remainder of this work. At this point, the coupler lifetimes are around 1.2~$\mu$s. At the chosen flux bias point, four distinct frequency transitions of the coupler are observed, see Fig.~\ref{fig:3}. A detailed description regarding the frequency selectivity is given in Appendix~\ref{sec:Design parameters}.
We verify that we can coherently drive conditional Rabi oscillations on the coupler transition by applying a microwave pulse to the driveline of $F_2$ which also capacitively couples to the transmon coupler. As shown in Fig.~\ref{fig:4}, depending on which state the fluxonium qubits are prepared in, we see a Rabi oscillation at different frequencies. Note that while the transmon coupler does not have an individual readout resonator, we can combine the readout signal of $F_1$ and $F_2$ to monitor the coupler dynamics, see also Appendix~\ref{Coupler readout}. Importantly, we notice that the fastest Rabi rate is observed when preparing the $\ket{11}$-state, thus, we expect that fastest possible gate to be achievable by driving this transition. 
Specifically, we apply a drive leading to the Hamiltonian:
\begin{align}
        \hat{H}_d (t) = [\hbar \Omega_I(t) \cos (\omega_d t) - \hbar \Omega_Q(t) \sin (\omega_d t)] \hat{n}_{F_2},
\end{align}
where $\Omega_I(t)$ and $\Omega_Q(t)$ are the pulse envelope of the in-phase and quadrature components of the drive, respectively and $\omega_d$ is the drive frequency. To achieve a fast high-fidelity gate, we must appropriately design the pulse shape of the drive as well as optimize the driving frequency. A black-box approach to finding the optimal driving parameters can be employed using the optimized randomized benchmarking for immediate tune-up (ORBIT) protocol~\cite{ORBIT} or using reinforcement learning algorithms~\cite{ding2023high}. However, the iterative feedback-loop between the experiment and the classical optimizer may lead to a time-costly calibration procedure. Here, we instead use a straightforward calibration approach with no iteration steps which allows for a fast and deterministic calibration procedure, see Appendix~\ref{tune-up}. In our gate scheme we choose to drive the population from the $\ket{101}$-state to the $\ket{111}$-state and back again to the $\ket{101}$-state to obtain the conditional phase (CZ) gate. To ensure that we recover all the population back to the $\ket{101}$-state, we first fix the pulse shape, i.e., $\Omega_I(t)$ and $\Omega_Q(t)$, and then we measure the coupler response as a function of the drive frequency and the absolute pulse amplitude of the drive. Thus, for each pulse frequency, we deterministically obtain the pulse amplitude that maximally recovers the population back into the $\ket{101}$-state, see also Appendix~\ref{tune-up}. Next, we measure the conditional phase from the application of 1, 3 or 5 gates. For a well-calibrated conditional phase gate, each of these should yield a conditional phase of exactly $\pi$. We measure the conditional phase for a range of frequencies and we pick the drive frequency where the conditional phase is equal to $\pi$. Finally, we measure the single qubit phases of both $F_1$ and $F_2$ subject to repeated application of the gate. We correct for these single qubit phases using virtual $Z$ gates~\cite{Virtual-z-gates}.

\begin{figure}[t]
\includegraphics[width=1\linewidth]{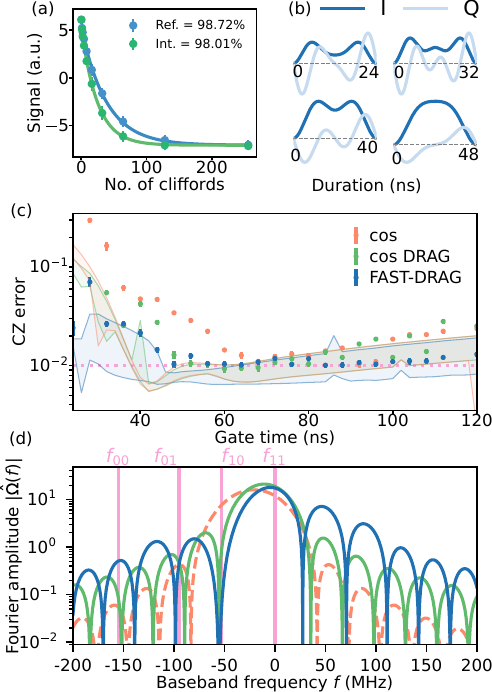}
    \caption{(a) Clifford interleaved randomized benchmarking for a two-qubit CZ gate. The readout signal for the reference sequence is shown in blue and in green we show the when interleaving the CZ gate. The extracted fidelities per Clifford is shown in the legend. (b) Pulse shapes used in the FAST-DRAG scheme are shown. Normalized I and Q components are shown in dark blue and light blue, respectively. 
    (c) Average CZ gate infidelity for the simple cosine pulse shape (orange), for the DRAG scheme (green) and for the FAST-DRAG pulse shape (blue). Error bars correspond to 1 standard deviation in the least-squares parameter fit. The horizontal dashed line  indicates a fidelity of 99\%. The shaded areas indicate the numerical simulations, see details in the main text. (d) Fourier spectrum with different pulse shapes is shown for a gate duration of 32~ns. Four vertical lines have been marked at the top corresponding to the transmon coupler frequencies $f_{00}$, $f_{01}$, $f_{10}$ and $f_{11}$.} 
    \label{fig:5}
\end{figure}

Using this calibration routine, we first explore a simple pulse shape where the in-phase drive is a raised cosine
\begin{align}
    \Omega_I(t) = \frac{\Omega}{2} [1 - \cos (2\pi t/t_g) ],
\end{align}
where $t_g$ is the total duration of the gate. Additionally, we set $\Omega_Q(t) = 0$. Using interleaved randomized benchmarking~\cite{Gambetta_RB,silva2024hands}, we characterize the gate fidelity of the CZ gate, see Fig.~\ref{fig:5}(a). For a cosine pulse, we find a maximal gate fidelity of $98.9\% \pm 0.1 \%$ for a gate duration of 68~ns. For longer gate times, the fidelity becomes limited by the coupler coherence times, see the extracted CZ errors in Fig.~\ref{fig:5}(c). For gate durations below 60~ns, the gate error increases significantly, which we interpret as resulting from driving other spurious transitions. A well-known technique to eliminate residual driving of weakly off-resonant transitions is the derivative removal by adiabatic gate (DRAG) technique~\cite{motzoi2009simple}. Using DRAG, we can eliminate the spurious driving at a detuning of $\Delta$ by defining the quadrature drive as:

\begin{equation}\label{eq:drag}
    \Omega_Q(t) = - \frac{1}{\Delta} \frac{d \Omega_I(t)}{dt}. 
\end{equation}

In our case the nearest transition that we are interested in suppressing is $\omega_{10}$ (see also Fig.~\ref{fig:3}). Thus, we can set $\Delta = \omega_{10} - \omega_d$ to obtain the pulse shape of the drive. Using this pulse shape, we see in Fig.~\ref{fig:5}(c) that we can obtain a gate fidelity around 99\% for a larger range of gate times compared to the simpler cosine pulse shape. However, the DRAG scheme can only be used to suppress one of the three residual transitions.  

To extend the degree to which we suppress the driving of residual transitions, we adopt the Fourier \textit{ansatz} spectrum tuning derivative removal by adiabatic gate (FAST-DRAG) method as introduced in Ref.~\cite{hyyppa2024reducing}. Specifically, we use a pulse envelope given by
\begin{align}\label{FAST-DRAG-eq}
    \Omega_I(t) = A \sum_n c_n \Big(1 - \cos \frac{2\pi n t}{t_g} \Big),
\end{align}
where $t_g$ is the gate duration, A is the overall amplitude and $c_n$ the coefficients for the cos series. Additionally, we add the quadrature drive following Eq.~\eqref{eq:drag} with the $\Delta =  \omega_{10} - \omega_d$ as before. As detailed in Appendix~\ref{sec:FAST}, we engineer the Fourier spectrum of the pulse by solving a convex minimization problem with three cos components in the series. In particular, we aim to minimize the Fourier components of the pulse in narrow frequency windows around $f_{01}$, $f_{10}$ and $f_{00}$. Examples of the resulting pulse shapes are displayed in Fig.~\ref{fig:5}(b) and the resulting average gate fidelities are shown in Fig.~\ref{fig:5}(c). We find a minimal gate error of $99.0\% \pm 0.1 \%$ for a gate time of 64~ns. It is interesting to note that while the overall lowest error is not improved compared to the DRAG scheme, the FAST-DRAG provides good performance over a broader range of gate duration showing that we more consistently accumulate less coherent errors from residual driving of spurious transitions. We visualize the impact of Fourier engineering by plotting the resulting pulse spectrum, see Fig.~\ref{fig:5}(d). Generally, we find that the simple cosine pulse has large frequency components around all spurious transmon transitions. In contrast, the DRAG scheme minimizes the component at $f_{10}$ as expected while the FAST-DRAG makes the window around $f_{10}$ broader while also suppressing the frequency components around $f_{01}$. Given the over-constrained optimization problem, we find that the frequency component of $f_{00}$ remains less suppressed.

To gain further insights into the errors, we simulate the gate numerically. We model the fluxonium qubits and the transmon coupler with the parameters discussed in Appendix~\ref{sec:Design parameters} and implement a time dependent drive using the \textrm{qiskit dynamics} package~\cite{Dynamics} . In the simulations, we include four energy-levels for each fluxonium qubit and three levels for the transmon coupler. From the resulting unitary evolution, we truncate the evolution to the computational subspace of the two fluxonium qubits to obtain the matrix $U$ from which we calculate the process fidelity as $F_p = |\text{Tr}(U_{ideal}^\dagger U)|^2/d^2$, where $d=4$ is the Hilbert space dimension. For each type of pulse shape and duration, we numerically optimize the drive amplitude and the drive frequency in order to maximize the gate fidelity. In the simulations, we include a drive on fluxonium qubit $F_2$ and residual relative driving on $F_1$ and on the transmon coupler. Specifically, we use a drive Hamiltonian $\hat{H}_D \propto \hat{n}_{F_2} + \eta_C \hat{n}_C + \eta_{F_1} \hat{n}_{F_1}$ to describe the relative capacitive crosstalk, $\eta_i$ from the driveline of $F_2$ to the transmon coupler $C$ and fluxonium qubit $F_1$. We fix $\eta_{F_1}=0.028$ based on capacitance matrix simulations of the device design. We find the simulated gate fidelities to be very sensitive to crosstalk, so to capture uncertainties in the precise crosstalk, we vary $\eta_{C}$ between 0.2 and 1.0.  The size of the full Hilbert space of two fluxonium qubits and the transmon coupler including the higher excited states prohibits efficient simulations as an open system. Thus, we include transmon $T_1$ in an ad-hoc manner. Specifically, we calculate the process fidelity of each transmon transition as
\begin{align}
    F_{p,ij} = \frac{1}{2}(1+e^{\int_0^{t_g} \gamma_{1,ij} p_{i1j}(t) dt}  ),
\end{align}
where $p_{i1j}(t)$ is the probability to be in the state $\ket{i1j}$ and where $T_{1,ij} = 1/\gamma_{1,ij}$ is the lifetime of the coupler when the fluxonium qubits are prepared in the state $\ket{ij}$. We calculate the total process fidelity as $F_{P,total} = F_P\prod_{ij} F_{ij}$ and finally we extract the average gate fidelity as $F_g = (dF_{P,total} + 1)/(d+1)$. As we increase $\eta_C$, the residual population in the transmon increases and, thus, the CZ errors increase as well. In Fig.~\ref{fig:5}(c), the shaded area indicate the span from the minimum to the maximal gate errors as we vary $\eta_{C}$ and we specifically observe that the cross-driving of the coupler can lead to an additional error of up to 1\%. For pulse durations longer than 60~ns, we see that the overall behavior of simulations matches well with the experimental data, thus, the coupler lifetime remains the main limiting factor for longer gate times. We also note that FAST-DRAG can potentially lead to smaller errors for longer gates since this drive scheme will cause less residual driving of the transmon and, thus, less potential errors due to the coupler lifetime. This effect is also partly visible in the experimental data. For the gate times smaller than 50~ns, we expect a lower error from the simulations, thus the experiment is clearly limited by residual driving of other transitions. From the simulations, we can identify residual driving of the higher level transitions of $F_1$ through the transmon coupler as the main cause of coherent errors. In fact, when there is large crosstalk to the transmon coupler, the simple cosine pulse is expected to outperform the FAST-DRAG scheme since the cosine pulse is overall more localized in frequency. Additionally, the simulations do not include crosstalk to the readout resonator detuned only 200~MHz from the drive frequency. For the short gate times, residual driving of the readout resonator may lead to additional dephasing of the qubits.

Finally, the simulations indicate that the simple cosine and the DRAG scheme should perform similarly. However, in the experiment we clearly see an improvement from using DRAG. The reason is likely that the numerical optimization scheme simultaneously optimizes the amplitude and detuning of the gate and finds a good compromise where a larger detuning is fine-tuned to get the precise conditional phase. In the experiments, we optimize the population recovery independent of the conditional phase, as described in Appendix~\ref{tune-up}. Thus, our simulations points to a clear direction for future devices, namely that we must minimize microwave crosstalk between the fluxonium qubits and improve the frequency targeting of the resonators and the coupler in a fluxonium-transmon-fluxonium architecture.

\section{\label{sec:conclusion}Conclusion}
In this work we have explored a system consisting of two fluxonium qubits coupled directly and through a transmon coupler circuit. The device parameters were designed such that the residual $ZZ$ crosstalk between the two fluxonium qubits is suppressed. In this architecture, we can selectively drive the transmon coupler based on the fluxonium qubits' state. Thus, by driving a full $2\pi$ rotation conditioned on the fluxonium qubits being in the $\ket{11}$ state, we were able to implement a conditional phase (CZ) gate. Using a simple cosine pulse shape for the drive, we obtained a fidelity of $98.9\% \pm 0.1 \%$ with a gate duration of 68~ns. By employing pulse shaping techniques, we optimized the performance further for shorter pulse durations. Specifically, we use the DRAG and the FAST-DRAG techniques to obtain fidelties around $99.0\% \pm 0.1\%$ over a larger range of gate durations. These pulse shaping techniques use simple analytical formulas to suppress the driving of residual coupler transitions and thereby minimizing the coherent errors from the drive.

\section{\label{sec:acknoledgements}Acknowledgments}
S.S., F.Y. and S.W. carried out the numerical simulations and calculations for the device energy parameters. S.S., F.Y. and P.K. fabricated the device. S.S., E.Y.H., J.H., M.F.S.Z. and T.V.S. characterized the device and performed the measurements for the experiments. S.S, E.Y.H., J.H., F.Y., M.F.S.Z. and T.V.S. contributed to the measurement setup. S.S wrote the manuscript with feedback from all coauthors. C.K.A. supervised the execution of the project and the writing of the manuscript. 

The authors thankfully acknowledge support from the Kavli Nanolab Delft cleanroom staff members including B.~van~Asten, E.J.M.~Straver, E.F.D.~Pot, B.P. van den Bulk, A.~van~Run, C.R. de Boer, M. Fischer, A.K.~van~Langen-Suurling  and M.~Zuiddam. We also acknowledge fruitful discussions with A.~Kamlapure, M.~Finkel, H.M.~Veen and D.J.~Thoen. The authors also acknowledge support from J.~Bauer and L.J.~Splitthoff on the junction development process.  

All authors acknowledge receiving support from the Dutch Research Council (NWO). E.Y.H. acknowledges funding support from Holland High Tech (TKI), J.H. acknowledges funding support from NWO Open Competition Science M and T.V.S. acknowledges support from the Engineering and Physical Sciences Research Council (EP-
SRC) under EP/SO23607/1.
\section{\label{sec:Data}Data availability}
The data for the experiments can be found at \cite{data_repo} with an open CC by 4.0 license. Scripts to analyze the data can be found in the gitlab repository at \cite{git_repo}   

\appendix 
\section{\label{sec:Design parameters}Device measurements}
\begin{figure}
    \centering
    \includegraphics[width=\linewidth]{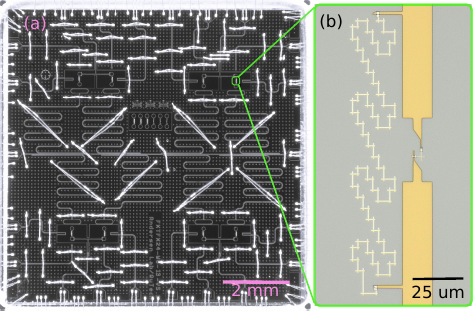}
    \caption{(a) The device is shown with all four sets of the three qubit system. The device can be seen with aluminum wirebonds on top to connect ground planes across control lines and Co-planar waveguide resonators. (b) A single fluxonium qubit zoomed in to show the array design and the single junction. The array comprises of total 100 junctions with a dimension of 450$\times$450 \text{nm}$^2$, and the single junction of 150$\times$150\text{nm}$^2$.}
    \label{fig:fab}
\end{figure}

\begin{figure}
    \centering
    \includegraphics[width=1\linewidth]{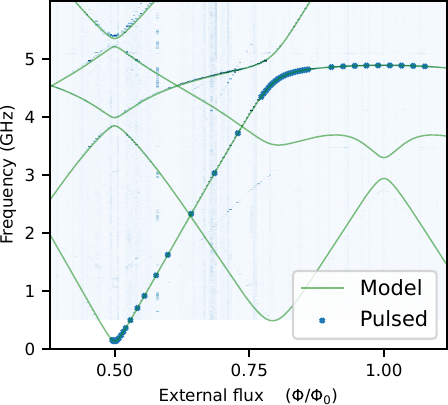}
    \caption{Two-tone continuous wave spectroscopy for fluxonium $F_2$ as a function of external flux and drive frequency shown in the background with spectroscopy peaks as blue crosses.
    A single fluxonium spectrum model is overlaid as solid green lines, with energy parameters fitted to the pulsed spectroscopy data.}
    \label{energy fit}
\end{figure}

In this section, we describe the device studied in this work and our approach to extract the device parameters. The device incorporates four pairs of fluxonium qubits each with a transmon coupler, as shown in Fig.~\ref{fig:fab}(a). In this work, we focus on a single set of fluxonium qubits.
Qubit spectroscopy for one of the fluxonium qubits is shown in Fig.~\ref{energy fit}, where pulsed spectroscopy peaks (blue markers) were fitted to a single fluxonium spectrum model. This fitting gives us the individual fluxonium energy parameters listed in Table~\ref{tab:Energies}.
The transmon capacitive charging energy is estimated by its measured anharmonicity and its Josephson energy is inferred from the measured coupler frequency. At the operational point, we measure the fluxonium qubits and the couplers lifetimes and coherence times using standard time-domain techniques, see Table~\ref{tab:Energies}.

\begin{table}[b]
    \centering
    \caption{Measured device parameters with F$_1$, F$_2$ being the fluxoniums and C being the coupler. The lifetimes of the coupler are measured with the fluxonium states prepared in $\ket{00},\ket{01},\ket{10}\&\ket{11}$,written in this order respectively. T$^E_2$ of the coupler for the $\ket{00}$ fluxonium state could not be measured, marked -. }
    \begin{tabular}{|c|c|c|c|} 
        \hline
        Measured & F$_1$ & C & F$_2$ \\ 
        \hline
       T$_1$ & 72.3$\mu$s & \makecell{0.54$\mu$s, 1.55$\mu$s\\1.05$\mu$s, 1.19$\mu$s} & 89.6$\mu$s \\
        \hline
        T$^R_2$ & 17.86$\mu$s & \makecell{0.77$\mu$s, 1.23$\mu$s\\1.19$\mu$s, 1.19$\mu$s} & 18.11$\mu$s \\ 
        \hline
        T$^E_2$ & 21.34$\mu$s & \makecell{-, 2.38$\mu$s\\2.45$\mu$s, 1.93$\mu$s} & 25.50$\mu$s \\ 
        \hline
        $\omega_{01}$ /2$\pi$ & 98.95 MHz & 4.58 GHz & 144 MHz \\ 
        \hline
        $\omega_{res}$ /2$\pi$ & 4.993 GHz & - & 5.082 GHz \\ 
        \hline
        E$_J$/h (GHz) & 4.9928 & 16.87 & 4.3350 \\ 
        \hline 
        E$_L$/h (GHz) & 0.5008 & - & 0.4921 \\ 
        \hline
        E$_C$/h (GHz) & 0.8805 & 0.1861 & 0.8829 \\ 
        \hline
    \end{tabular}
    \label{tab:Energies}
\end{table}

\begin{figure*}[t]
    \centering
    \includegraphics[width=1\textwidth]{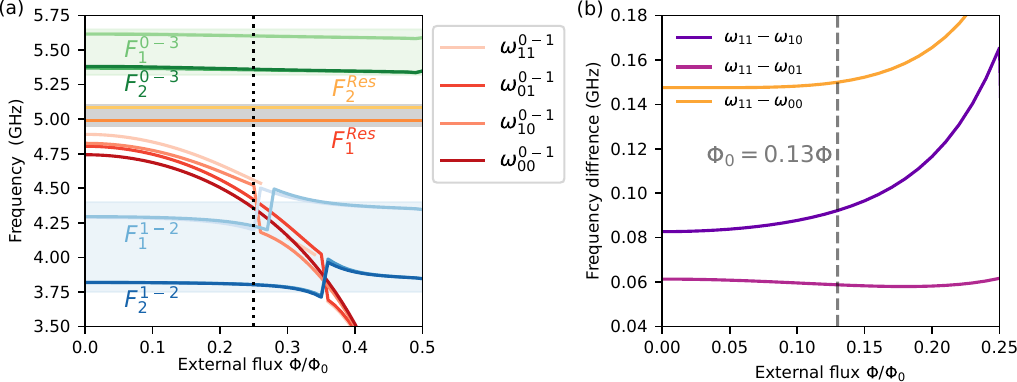}
    \caption{(a) Energy levels of the fluxonium-transmon-fluxonium system as a function of the external flux of the transmon. The transition frequency from state $\ket{n}$ to $\ket{m}$ for fluxonium $i$ is indicated with $F_i^{n-m}$. In the blue and green regions we see the 12 and 03 transitions of the fluxoniums respectively. The grey region contains the two readout resonators. The coupler transitions are labeled $\omega_{ij}^{0-1}$ corresponding to the fluxonium qubits in state $\ket{ij}$.  (b) Differences between the undesired transition of the transmon (with the fluxoniums in $\ket{00},\ket{01},\ket{10}$) and the desired transition $\ket{11}$ as a function of the external flux of the transmon.. $\omega_{ij}$ represent $F_1$ prepared in the $i$th state and $F_2$ in $j$th state, respectively. A black vertical dashed line is marked to show value of $\Phi = 0.13 \Phi_0$ corresponding to the operational point used in the main text.    }
    \label{fig:energy_diag}
\end{figure*}

We use the the \textrm{scqubits} package~\cite{scqubitsDocumentation} to analyze the energy spectrum of the system in more detail. In Fig.~\ref{fig:energy_diag}(a), we see how the transmon transition frequencies are positioned between the $\ket{1}$ to $\ket{2}$ transitions and the $\ket{0}$ to $\ket{3}$ transitions of each fluxonium. In particular, we notice a strong avoided crossing between the $\ket{0}$ to $\ket{1}$ transition of $C$ and the $\ket{1}$ to $\ket{2}$ transition of $F_1$ as we increase the external flux bias of the transmon coupler. Moreover, it is essential to ensure that the coupler is sufficiently detuned from the fluxonium resonators to avoid Purcell decay of the coupler transmon into the readout resonator. At an external flux of $\Phi/\Phi_0=0$, the transmon coupler transitions approaches the readout resonator frequencies. Biasing the transmon coupler away from zero external flux reduces the Purcell decay rate, although the coupler becomes more sensitive to flux noise. 

Furthermore, as seen in Fig.~\ref{fig:energy_diag}(b), when we bias the transmon coupler away from zero external flux, we also increase the frequency selectivity of the desired transitions frequency $\omega_{11}$ relative to the $\omega_{00}$ and $\omega_{10}$.

\section{\label{sec:fabrication}Fabrication}
The device used in this work has been fabricated using a recipe similar to those described in \cite{yilmaz2024energy,stefanski2024improved}, we will here describe further details as several improvements have been implemented in the process. The process begins with cleaning a high-resistivity silicon wafer (\(>\)20 k\(\Omega\)·cm) with a (100) orientation, sourced from Topsil. The wafer is first immersed in a nitric acid bath for 7 minutes, accompanied by sonnication, to remove organic residues from the surface. Next, it is sequentially rinsed in two deionized water baths for 60 seconds each before being dried with nitrogen gun. The wafer undergoes an oxide removal step using a 40\% HF solution for 6 minutes, followed by another deionized water rinse as described above. To further passivate the surface and delay oxide formation, the wafer is coated with HMDS vapor~\cite{bruno2015reducing}. Since this passivation has a limited effective duration, the wafer is immediately loaded into a sputter system. For deposition of NbTiN, we utilized a sputtering system with a confocal target geometry. The deposition is performed using an NbTi target with a chemical composition of Nb:Ti (70:30). 
We deposit 200~nm of NbTiN using a deposition rate of 28.3~nm/s.
To further process the wafer, we dice it into smaller pieces of $18\times 18$ $mm^2$. 
To remove the protective resist from the dicing step, the device is put in NMP for 2 hours at 80~$^{\circ}$C and additionally left in NMP at room temperature over night. 
To remove potential resist residues, we follow the overnight cleaning step with IPA for a 1 min, a 65~$^{\circ}$C PRS for 20 minutes and finally IPA for 1 min.

The next step is to pattern the NbTiN using electron beam lithography. To prepare the sample for the lithography step, an anisotropic oxygen plasma descum, with 20sccm $O_2$ at 0.1mbar and 20W for 2 minutes, is used to remove organic residues. To improve resist adhesion, we bake the sample for 10 minutes at 150$^{\circ}$C and spin a HMDS layer. We use AR.P6200.18 as resist, which we spin at 2500 rpm and bake at 160$^{\circ}$C for 3 min. We finally pattern the coplanar waveguide and capacitive structures of the device using a Raith 5200 EBPG system using a beam current of 250~nA and a dose of 350 $\mu$C/cm$^2$. After exposure, the resist os developed in a multi step process: (i) Amyl acetate development for 30~s with sonnication, (ii) Amyl acetate development for 30 s without sonnication in a separate beaker, (iii) MIBK for 10 s, (iv) IPA for 60 s, (v) MIBK for 20 s with sonnication, (vi) 2 times: IPA for 15 s in separate beakers. To remove resist residues an additional 2-minutes oxygen descum is performed as in the previous descum step. The pattern is etched using reactive ion etching system in a two-step recipe: The first step uses $SF_6$:$O_2$ 13.5:4 at 70 W and a pressure of 0.1~mbar until a signal-drop from the end-point detection system. The second step step uses  $SF_6$:$O_2$ 4: 16 at 50 W and a pressure of 0.08~mbar. We perform a final descum step with a 220 sccm flow of $O_2$ at 200~W for 30 s. 

To remove the resist, we use the resist remover PRS-3000 first for 1 minute with sonnication followed by 3 hours without sonnication in a separate beaker.
To further remove any organic residues from the surface, we clean for 30~s  using nitric acid, followed by 25 minutes cleaning using 7: 1 BOE to remove any native oxides \cite{altoe2022localization}.

The Josephson junctions are patterned with E-beam lithography. This process uses a trilayer stack of MMA El8; PMMA 495 A8 and PMMA 950 A3 which we spin at 3000, 1000 and 3000 rpm, respectively. Each spinning step involves baking at 185$^{\circ}$C for 10~minutes and cooling down for 2~minutes. The junctions are defined by exposing the resist using a 5nm resolution beam and a current of 1nA. The exposure is designed to define a tapered undercut profile~\cite{muthusubramanian2024wafer} using a selective exposure of the MMA resist. The undercuts are defined using a dose of 170 $\mu$C/cm$^2$ and we use a ratio of the junction dose to the undercut dose of 6:1. The development of the resist is performed in a H2O:IPA (1:3) mixture at room temperature with sonnication for 3 min 10s. Subsequently, the device is cleaned in IPA 2 times for 10~s and 1~minute, respectively.
To clean resist residues, we do an oxygen descum, followed by cleaning with a 7:1 BOE mixture~\cite{zanuz2024}. Immediately afterwards, we load the device into a Plassys MEB550 system for the electron-beam evaporation process. 
The junction deposition starts with 30~nm Al deposition followed by a 5~mbar 20 min oxidation step. Then we deposit 110~nm of Al after rotating the sample 90$^\circ$ followed with another oxidation at 1.3~mbar for 11 mins. 

The chip is cleaned using NMP at 80$^{\circ}$C for 2 hours and then left overnight at room temperature. Next, the device is cleaned again using NMP while removing the residual aluminum film with a pipette. We additionally clean the device in 2 more times with NMP and 3 additional times with IPA for 5 minutes each.

To create a better galvanic connection between the junctions and the base layer, we deposit additional aluminum bandages~\cite{dunsworth2017characterization}. We spin a resist stack of MMA EL8 and PMMA 950 A4 at 4000 and 2000 rpm, respectively. 
Each resist layer is baked at 180$^{\circ}$C for 5 minutes. 
The resist is exposed using electron-beam lithography with a beam current of 250~nA and a dose of 1200 $\mu$C/cm$^2$ and developed with s H2O:IPA (1:3) mixture at 6$^{\circ}$C for 1 min 30 s followed by 30 s IPA.

Prior to the Al~deposition a ion-milling step is performed to remove native oxides and resist residues and we deposit 150nm of Al film followed by 1.3-mbar 11 min oxidation. The lift-off is done similar to the junction lift-off step. 
The final result is illustrated in Fig~\ref{fig:fab}(b), where we see both the smaller Josephson junction as well as the larger Junctions in the array that were fabricated simultaneously in this process. The whole device is finally diced again and we wirebond the device to a PCB as shown in Fig~\ref{fig:fab}(a).

\begin{figure}
    \centering
    \includegraphics[width=\linewidth]{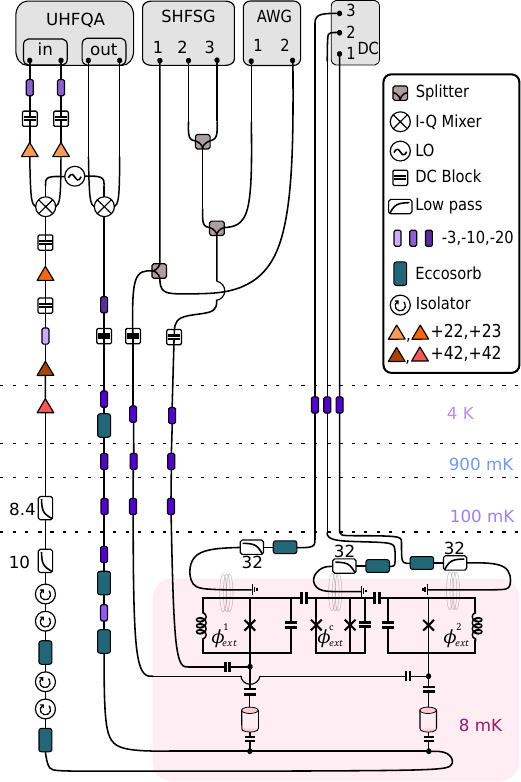}
    \caption{Schematic of the wiring in the dilution fridge and electronic setup used during the experiment. Inset shows different components, the numbers shown represent the attenuation and amplification in dB. Filters used on the output line of the transmission chain are written in GHz, while the filters connected on flux lines are represented in MHz.     }
    \label{fig:wiring}
\end{figure}

\section{\label{sec:citeref}Experimental setup}
The PCB holding the device is attached to a gold-plated copper mount and covered with two rectangular aluminum shields.
The assembly is installed in a Blue Fors LD-400 system. The sample holder is further covered with three coaxial shields, consisting of two mu-metal cans and one gold-plated copper can. The PCB is connected to room-temperature electronics with coaxial microwave cables as shown in Fig.~\ref{fig:wiring}.

We use a Zurich instruments (ZI) UHFQA to generate the input and output signals for readout. The signals are up- and down-converted using a ZI HDIQ IQ modulator and a custom built IQ modular, respectively. The LO is provided by a signal generator from from Anapico. The output signal is amplified at the 4K stage using a HEMT amplifier from Low-Noise factory and at room temperature using a series of amplifiers as shown in Fig.~\ref{fig:wiring}.
For driving the individual fluxonium qubits, we a ZI HDAWG which generates signals directly at the qubit frequencies. Additionally, we use a ZI SHFSQ to drive the transmon coupler as well as for qubit reset, see also Appendix~\ref{reset}. These signals are combined with power combiners at room temperature. 
For flux bias of the qubits, we use in-house built current source modules (S4G). The flux lines on each qubit have a 32~MHz low pass filters to filter out noise at the fluxonium qubit frequency. 
\section{\label{zz}ZZ measurement}
\begin{figure}[t]
    \centering
    \includegraphics[width=1\linewidth]{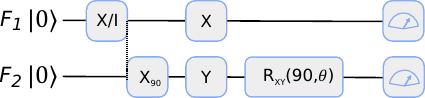}
    \caption{
    Circuit for measuring the residual $ZZ$ interaction rate.
    }
    \label{fig:zz}
\end{figure}

The residual $ZZ$ rate, as defined in Eq.~\eqref{eq:residual-zz-definition}, was measured using the circuit in Fig.~\ref{fig:zz}.
One fluxonium is designated as the control qubit and is initialized in either the $|0\rangle$ or $|1\rangle$ state. The other fluxonium is initialized in an equal superposition state.
Both qubits freely evolve for a duration $\tau/2$ and we then apply refocusing $\pi$ pulses on both qubit simultaneously.
Both qubits then freely evolve again for a duration $\tau/2$. Finally, a $\pi/2$ recovery rotation is applying for a set of different rotation angles $\theta$. During the free evolution time, the target qubit rotates around the $Z$-axis by $+\chi_{ZZ}\tau/2$ ($-\chi_{ZZ}\tau/2$) when the control qubit is prepared in the $|0\rangle$ ($|1\rangle$) state. This phase offset can be extracted from a cosine fit to the target qubit population as a function of $\theta$.

\begin{figure}
    \centering
    \includegraphics[width=1\linewidth]{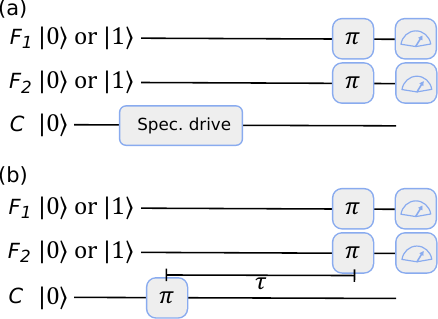}
    \caption{(a) Circuit to measure the conditional spectroscopy for the coupler and fluxoniums is shown. (b) Circuit to measure the coherence times of the coupler. }
    \label{fig:coupler}
\end{figure}

\section{\label{Coupler readout}Coupler readout and Rabi}
In this work, we perform readout of the coupler using a Rabi assisted readout scheme~\cite{zhangcharacterization} since the coupler does not feature a dedicated readout resonator. 
In the scheme, the state of the transmon is mapped onto the fluxonium qubits using $\pi$-pulses resonant with the $|000\rangle\to|100\rangle$ and $|000\rangle\to|001\rangle$ transitions. In this way, we can perform spectroscopy of the coupler by simply appending this readout scheme to a spectroscopy pulse at the coupler frequency, see Fig.~\ref{fig:coupler}(a). Given the chosen resonance condition for the $\pi
$-pulses, the fluxonium qubits will only change state if the transmon coupler is in its excited state. i.e., if the spectroscopy pulse was off-resonant.
Subsequently, by performing principal component analysis on the two transmission amplitudes, we extract a single measurement signal from the main principal component.

To characterize the lifetime of the coupler, we use the circuit shown in Figure~\ref{fig:coupler}(b), where the coupler is excited with a calibrated $\pi$ pulse. In this measurement, we initialize each fluxonium qubit in either $\ket{0}$ or $\ket{1}$ to extract any state-dependent lifetime of the coupler. 

\begin{figure}[t]
    \centering
    \includegraphics[width=\linewidth]{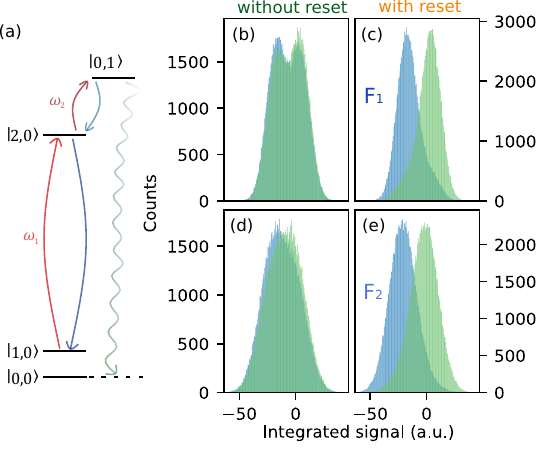}
    \caption{(a) Reset scheme used for initializing the qubit in the ground state. The energy levels are labeled with $ |Q,R\rangle$ where Q is the qubit state and R is the resonator state shown with simultaneous driving at the angular frequencies $\omega_1$ and $\omega_2$ and with relaxation back into the qubit ground state. (b),(c) Single shot readout histograms for qubit $F_1$ prepared in $|0\rangle $ in (b)  and $\ket{1}$ in (c) displayed in blue and green, respectively. Panel (b) is the outcome without performing the reset, while (c) is the result after we apply the reset. Similarly panel (d),(e) represent data for qubit F$_2$. }
    \label{fig:reset}
\end{figure}
\section{\label{reset}Reset}

\begin{figure*}[t]
    \centering
    \includegraphics[width=1\textwidth]{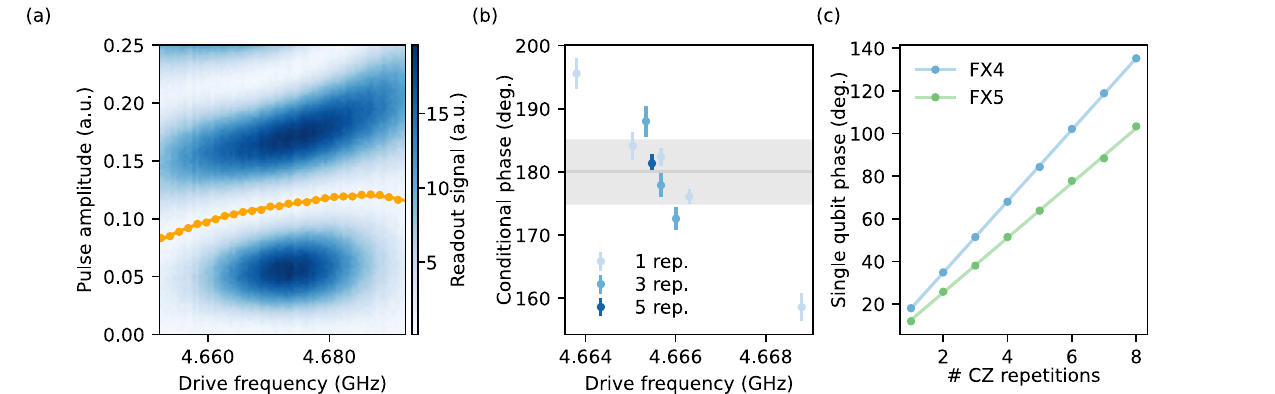}
    \caption{
    (a)
    Coupler response to varying drive frequency and pulse amplitude.
    The heatmap corresponds to the magnitude of the difference in the Rabi assisted readout signal when a drive is or is not played, see main text.
    The yellow markers indicate the $2\pi$ rotation amplitude for each drive frequency and the solid yellow line is a spline fit through the markers.
    (b) Conditional phase vs. drive frequency with varying gate repetitions. The gray band represents a conditional phase error of $\pm5$ degrees.
    (c) Single qubit phase errors amplified with increasing number of gate repetitions. }
    \label{fig:CZ-tune}
\end{figure*}

We reset the qubits using a two-tone driving scheme similar to~\cite{Fluxonium_reset}. 
The qubit reset protocol uses two drive pulses at the angular frequencies $\omega_1$ and $\omega_2$. The drive frequencies correspond to transition frequency between $|1,0\rangle$ and $|2,0\rangle$ and to the transition frequency between $|2,0\rangle$ and $|0,1\rangle$, respectively. 
When the system is in the state $\ket{0,1}$, it will decay back into $|0,0\rangle$ by the dissipation of one photon through the resonator, see Figure~\ref{fig:reset}(a). The reset uses a square bichromatic drive with a duration of 32~$\mu$s followed by an idle time of 8~$\mu$s to ensure that the resonator is fully dissipated. 
When the qubits (F$_1$ and F$_2$) are not reset, they thermalize into a mixed state with almost equal populations in $\ket{0}$ and $\ket{1}$. As a result, the single shot readout histogram does not clearly distinguish between the qubit states, see Fig.~\ref{fig:reset}(b) and (d). After applying the reset, the qubit is initialized in $\ket{0}$ states and we now clearly see two well-separated histogram peaks in Fig.~\ref{fig:reset}(c) and (e). 
\section{\label{tune-up}CZ gate tune-up}
In this work, to calibrate the gate, we first measure Rabi oscillations on the coupler when the fluxoniums are prepared in the $|11\rangle$ state, see Fig.~\ref{fig:CZ-tune}(a) where the pulse duration and pulse shape is kept fixed. From this measurement, we extract the pulse amplitude for each drive frequency that maximally returns the transmon back to its initial state, see orange markers in Fig.~\ref{fig:CZ-tune}(a) which we fit with a spline function to establish a map from drive frequency to drive amplitude.

Using the extracted drive amplitudes, we determine the conditional phase from the CZ gate as a function of the drive frequency, see Fig~\ref{fig:CZ-tune}(b). Specifically, we perform a measurement of the conditional phase for 1, 3 and 5 gate repetitions. To efficiently sample the frequency axis, we use a noise aware root finding algorithm to locate the point of $\pi$ phase. Once the phase error is within $\pm5$ degrees, we increase the number of gate repetitions and repeat the procedure and we terminate when 5 gate repetitions yield a phase error less than 5 degrees.

Once the pulse amplitude and frequency is calibrated to generate the conditional phase gate,
we calibrate the residual single qubit phases for both fluxonium qubits.
The single qubit phases are measured by amplifying the single qubit phase errors with repeated applications of the CZ gate, see Fig.~\ref{fig:CZ-tune}(c).
We linearly fit the accumulated phase data and use the slope as a measure of the acquired phase per gate. The unwanted single qubit phases are compensated by using virtual-$Z$ gates
\cite{Virtual-z-gates}.

\section{\label{sec:FAST}FAST-DRAG calculations}
To eliminate the undesired frequency components in the in-phase pulse envelope ($\Omega_I$) of the drive used for the gate, we implemented the FAST-DRAG method~\cite{hyyppa2024reducing} to derive analytical solutions for the shape of the pulse. This approach works by attenuating the frequency amplitude across $n$ frequency spans $(f_n)$ each spanning a range from $f_{n,l}$ to $f_{n,h}$. Additionally, the amplitude of each frequency span is suppressed by an amount proportional to a specified weight ($w_n$) assigned to that span. 
The method works by defining the pulse envelope, $\Omega_I$, as a cosine series as described in Eq.~\eqref{FAST-DRAG-eq}
\begin{equation}
    \Omega_I =  A \sum_{n=1}^{N} c_n g_n(t),  
\end{equation}
where g$_n$= $(1-\cos\frac{2\pi nt}{t_g})$ is cosine pulse with $n$ period for the duration of $t_g$. The pulse amplitude is given by $A$ and each cosine component is weighted with a coefficient $c_n$. To solve for the coefficients $c_n$ such that the combined pulse shape minimizes the frequency components at the undesired frequencies, we choose $N=3$ basis functions and solve the following quadratic optimization problem
\begin{align}\label{eq.min}
    \text{minimize} \sum_{n=1}^{3}w_n\int_{f_{n,l}}^{f_{n,h}}|\hat{\Omega}_I(f)|^2\,df,\\
    s.t \sum_{n=1}^{3}c_nt_g=\pi,
\end{align}
where $\hat{\Omega}_I(f)$ is the Fourier transform of the pulse envelope. The condition of $\pi$ comes from the desired angle of rotation for the gate. To provide an experimental basis to the mathematical relations, we define the center frequencies of each frequency span to be the conditional transitions of the coupler $\{f_{00},f_{01},f_{10}\}$ not corresponding to the $|11\rangle$ state of the fluxonioums. Additionally, we assign the weights of each spurious frequency as the square of the Rabi rate of the coupler, corresponding to the transition, see Fig.~\ref{fig:4}. Further, we use a frequency span of 5~MHz centered at each transition. 

To find the appropriate minimum as given in Eq.~\eqref{eq.min}, we first consider the Fourier transform of basis function given by
\begin{equation}
\begin{split}
    \hat{g}_n
    ={}&
    t_g(e^{-\text{i}\pi t_gf}\text{sinc}(\pi t_gf)
    \\
    &-
    \frac{1}{2}e^{\text{i}\pi (n/t_g -f) t_g}\text{sinc}(\pi (n/t_g -f) t_g)
    \\
    &+
    \frac{1}{2}e^{\text{i}\pi (n/t_g +f) t_g}\text{sinc}(\pi (n/t_g +f) t_g)
\end{split}
\end{equation}
such that we now need to minimize
\begin{equation}\label{eq.minfunc}
\sum_{n=1}^{3}w_n\int_{f_{l,n}}^{f_{h,n}}\Big|\sum_{i=1}^{3}c_n\hat{g}_n(f)\Big|^2.
\end{equation}
We can re-write Eq.~\eqref{eq.minfunc} as
\begin{equation}
    \sum_{n,m}c_nc_mA_{nm} = c^TAc
\end{equation}
where A$_{nm}$= $\sum_{i=1}^{3}w_i\int_{f_{i,l}}^{f_{i,h}}\hat{g}_n\text{(f)}\hat{g}_n^*\text{(f)}df$ and $c$ is a vector containing the coefficients $c_n$. The minimization problem now become equivalent to solving the matrix equation 
\begin{equation}
    \begin{bmatrix}
A + A^T & -b \\
b^T & 0 
    \end{bmatrix} \times (c^T,\mu)^T = (0^T,\pi)^T
\end{equation}
where b = (1,.....1)$^T \in \Re^{N\times 1}$, where we have introduced the Lagrange multiplier $\mu$. The pulse shapes presented in the main text are found using the solutions to this matrix equation. 

\bibliography{apssamp}
\end{document}